# Dynamics of the Electo-Reflective Response of TaS$_3$

R.C. Rai and J.W. Brill

*Department of Physics and Astronomy, University of Kentucky, Lexington, KY*

*40506-0055*

Abstract

We have observed a large (~1%) change in infrared reflectance of the charge-density-wave (CDW) conductor, orthorhombic TaS$_3$, when its CDW is depinned. The change is concentrated near one current contact. Assuming that the change in reflectance is proportional to the degree of CDW polarization, we have studied the dynamics of CDW repolarization through position dependent measurements of the variation of the electro-reflectance with the frequency of square wave voltages applied to the sample, and have found that the response could be characterized as a damped harmonic oscillator with a distribution of relaxation (i.e. damping) times. The average relaxation time, which increases away from the contacts, varies with applied voltage as $\tau_0 \propto 1/V^p$ with p ~ 3/2, but the distribution of times broadens as the voltage approaches the depinning threshold. Very low resonant frequencies (~ 1 kHz) indicate a surprisingly large amount of inertia, which is observable in the time dependence of the change in reflectance as a polarity dependent delay of ~ 100 μs.

Interest in quasi-one dimensional conductors with sliding charge-density-waves (CDW's)[1,2] has continued for almost three decades because of the variety of unusual properties they exhibit. In the CDW ground state, a periodic lattice distortion is accompanied by a modulated electron density: $n = n_0 + n_1 \cos[Qz + \varphi(z,t)]$, where Q is the CDW wavevector, $n_1$ its amplitude, and z is the direction of conducting chains. Because CDW pinning results from the competition between deformations of the CDW (i.e. variations of $n_1$ and $\varphi$) and its interactions with impurities, CDW's are also model systems for studying the effects of quenched disorder on a deformable periodic medium.

When a voltage greater than its depinning threshold is applied, the CDW can slide through the sample, carrying current.[1] At the same time, the CDW becomes elastically strained; i.e. its phase varies throughout the sample so that the CDW is compressed near one current contact and rarefied at the other [3-5]. The strain ($Q^{-1}\partial\varphi/\partial z$) near a current contact is required to drive the phase-slip process needed for current conversion, i.e. to allow electrons to enter and leave the CDW condensate at the contacts,[3,6] while the smaller strain near the center of the sample reflects the shift in the chemical potential due to imperfect screening by the uncondensed (or thermally excited) quasiparticles.[4] The latter polarization can be frozen into sample for long times when the current is removed, as its relaxation requires the climb of CDW phase dislocations.[4] Studies of the dynamics of CDW polarization from an unpolarized state[7] have hence been hindered by this metastable behavior in the pinned state.

On the other hand, if the applied voltage is reversed above threshold, the CDW strain also (approximately) reverses[3,6] and the CDW changes between two states of dynamic equilibrium. Hence, transport measurements with an ac amplitude larger than threshold could, in principle, be used to study the dynamics of CDW "repolarization". In NbSe$_3$, which stays metallic in its CDW state,[1] this repolarization time was found to be surprisingly long (~ 10 µs)[3] and inductive behavior was observed for frequencies near 1 MHz.[8] Longer times are expected in semiconducting CDW materials, but we are not aware of any such measurements. However, in semiconducting blue bronze, K$_{0.3}$MoO$_3$,[1] the initial CDW polarization (i.e. from an unpolarized state) was found to obey a stretched exponential time dependence; the characteristic polarization time, presumably ~ 50% of the repolarization time, was found to be an activated function of voltage, $\tau \propto \exp V_0/V$, with a typical time constant ~ 0.1 ms for temperatures near 80 K and voltages slightly above threshold.[7]

In measurements of the infrared transmittance of blue bronze, we found that, for photon energies below the CDW gap, the transmittance[5] varied spatially in a sample when a voltage above threshold is applied, with maximum changes (~ 1%) adjacent to the contacts. The spatial dependence suggested that the relative change in transmittance was proportional to the CDW strain, and we initially suggested that the transmission changes resulted from changes in the intraband absorption of quasiparticles, whose density changed to screen the strains.[5,9] In subsequent measurements, we found that the phonon spectrum was also affected by the applied voltage,[10] but these changes also seemed to be proportional to the CDW strain. Hence if an alternating square wave voltage, ±V(Ω) was applied to the sample, the frequency (Ω), voltage, and position dependence of the CDW repolarization could be studied through measurement of changes in the optical properties. Indeed, the characteristic repolarization time we observed,[5] ~ 0.1 ms was similar to the pristine polarization time observed in Reference [7], but detailed measurements of the frequency dependence of the electro-transmittance of blue bronze were not made.

In subsequent work,[11] we showed that the reflectance (R) of blue bronze varied with voltage generally as expected from its electro-transmittance. In particular, the electro-reflectance spectra were dominated by phonon, rather than quasiparticle, changes and consequently the changes were small, with relative changes in reflectance, $|\Delta R/R| \sim 0.1\%$ for transversely polarized light and a few times smaller for parallel polarized. Consequently, the sign of the reflectance change (at a given position and voltage) varied as the wavelength was tuned through phonon lines. However, the dependence of $\Delta R/R$ (for a given wavelength) on position was slightly different than for the electro-transmittance, in that the magnitude of the electro-reflectance signal decreased (and, for low voltages and frequencies, the signal became inverted) near the contacts; these differences suggested that the CDW strain was affected by the sample surface, perhaps due to the placement of the contacts.

In this paper, we discuss the electro-reflectance of the related semiconducting CDW compound, orthorhombic $TaS_3$;[12] this is the first report of a position/voltage dependent change in the optical properties of this material.[13] We have observed that, for polarizations both parallel and transverse to the conducting chains, $|\Delta R/R| \sim 1\%$, an order of magnitude larger than for blue bronze, with typical time constants also a few times larger than for blue bronze.[11] We have studied the voltage-frequency-position dependence of the electro-reflectance signal, and, *assuming that the response is proportional to the local CDW strain*, discuss its implications on the dynamics of CDW repolarization.

Orthorhombic TaS$_3$ undergoes a CDW transition into a semiconducting state at $T_c$ = 220 K.[1,12] As grown crystals, with typical dimensions 2 x 0.02 x 0.005 mm$^3$, were cleaned with acetone and alcohol and mounted with silver paint to span a 1 mm hole in a sapphire ring. Gold films were evaporated at the ends of the sample to serve as current contacts, with the films extending slightly into the hole so that the effective length of the sample between the films was ~ 0.8 mm. The samples were placed in a commercial infrared microscope[11] operating in a reflectance mode. Tunable diode lasers[10,11] were used as light sources. For the measurements discussed in this paper, the light spot was focused to a spot ~ 50 μm long along the length of the sample and slightly wider than the sample; light that missed the sides of the crystal went into the hole in the substrate and therefore was not reflected back to the detector.

Symmetric bipolar square wave voltages (±V) were applied to the sample while the light was chopped at an incommensurate frequency. The response of the MCT detector in the microscope was measured with two lock-in amplifiers operating at the square wave (Ω) and chopping (ω) frequencies; the relative change in reflectance is given by ratio of the lock-in signals, S:[11]

$$S(\Omega)/S(\omega) = \Delta R/R \equiv [R(+V) - R(-V)]/R_{ave}, \quad (1)$$

where $R_{ave}$ is the average reflectance. For these dynamics studies, the in-phase and quadrature components of S(Ω) were measured separately, requiring determination of the frequency dependence of the electronics. The phase shift of the microscope detector amplifier could only be determined to a precision of ~ 2°; this precision was adequate at most frequencies, but led to relatively large uncertainties in the quadrature component of the electro-reflectance signal at low frequencies (e.g. < 100 Hz) where the quadrature component is much smaller than the in-phase component (e.g. see figure 2, below).

All measurements were made with the sample holder at T ~ 77 K; the sample, in vacuum, may have been a few degrees warmer than this in general, and an additional few degrees warmer for measurement at high voltage. In principle, since measurements were made with symmetric square wave voltages, there should not have been an oscillating signal due to Joule heating of the sample, but CDW materials often have slightly unequal responses to positive and negative currents, so there may have been a small thermal signal for frequencies below the thermal response rate (~ 100 Hz). However, since the electro-reflectance response for bipolar square waves was always much larger than we observed for unipolar square waves (e.g. see Reference [11] for similar results in blue bronze), the thermal component to the signals is considered negligible.

Although samples were chosen for having visually flat surfaces, we typically found that the reflectance signal varied considerably as we scanned the light along the sample, suggesting that the reflectance was affected by unseen steps and/or facets (perhaps causing diffraction as well as misdirecting the light). Similarly, we were not able to take meaningful spectra with our lasers. However, these affects should not affect the relative change in reflectance, as determined by Eqtn. (1). When studying a new sample, we scanned the spectrum for wavelengths at which the electro-reflectance could be measured with a high signal/noise ratio. (As discussed in References [10,11], there was considerable, variable noise due to mechanical vibrations of the helium refrigerator cooling the lasers). All the data shown below is for parallel polarized light, but no significant differences (with caveats noted later) were observed for transversely polarized light.

The inset to Figure 1 shows the square-wave voltage dependence at $\Omega/2\pi = 253$ Hz of $\Delta R/R$ for sample 1, measured with light of energy $\nu = 860$ cm$^{-1}$ focused adjacent to a current contact. (Note, as indicated by the minus sign on the ordinate, also for Figures 2 and 4, that at this wavelength the reflectance decreases near the positive contact.) The onset voltage of the electro-optical signal, $V_{onset} = 40$ mV, is presumably the threshold at which the CDW becomes depinned within the crystal.[9] The threshold for non-linear current is higher (i.e. $V_T = 70$ mV), and is the voltage at which the CDW is depinned at the contacts; the difference is the "phase-slip voltage".[3,6]

The spatial dependence at 253 Hz at a few voltages is shown in the main part of Figure 1. Note that, unlike for blue bronze, the spatial profile is very asymmetric, with the response much larger on one side of the sample. Similar asymmetry was observed for all six samples studied, although for one sample the response was quite symmetric on its initial cool down and only became asymmetric with subsequent coolings. This result suggests that the CDW polarization, either because of differences in the electrical contacts or defects in the sample, tends to pile up on one side of the sample. Note that the electro-reflectance signal is a very irregular function of position on the other side of the sample (i.e. for $x > 250$ μm), and even again changes sign near the $x \sim 800$ μm contact for small voltages; this inversion is similar to what is observed for blue bronze,[11] as mentioned above.

Figure 2 shows the frequency dependence at $x = 0$ for this sample at a few voltages, for frequencies between 25 Hz and 4 kHz. As expected, the average relaxation time, estimated from the reciprocal of the frequency of the peak in the quadrature response (and shoulder in the in-phase response) decreases as the voltage increases.[7] An unexpected feature, however, is that the in-phase response becomes inverted for frequencies greater than ~ 1 kHz, indicative of surprisingly large inertia to repolarization. In contrast, the inductive response in NbSe$_3$, corresponding to the inverted in-phase response, was only observed for frequencies near 1 MHz.[8] The curves show fits to the modified damped oscillator equation:

$$\Delta R/R = \Delta R/R)_0 / [1 - (\Omega/\Omega_0)^2 + (i\Omega\tau_0)^\gamma]. \quad (2)$$

In analogy to the Cole-Cole generalization of simple relaxation,[14] we've introduced the exponent $\gamma$; values of $\gamma < 1$ imply a distribution of relaxation times.[15]

The voltage dependence of the parameters of the fit, for light at two locations each for two samples, are shown in Figure 3. In each case, $\tau_0 \; \alpha \; V^{-p}$, with $1 \leq p \leq 3/2$, much weaker than the exponential voltage dependence in blue bronze found from transport measurements mentioned above.[7] There is no indication of $\tau_0$ diverging at the depinning threshold, as expected for a dynamic critical models of depinning,[16] but while data was taken down to the nonlinear threshold, $V_T$, we have not yet been able to approach the "bulk threshold", $V_{onset}$. However, what is observed in most cases is that the exponent $\gamma$ decreases at small voltages, indicating a broadening of the distribution of relaxation times. While there is no clear voltage dependence to the resonant frequency, its typical value of $\Omega_0/2\pi$ ~ 1 kHz suggests that there is ~ 100 μs delay between the applied voltage and the electro-reflective response.

For both samples, the relaxation time 100 μm from the contact is greater than at the contact. This is consistent with previous results[3,5] and suggests that the relaxation is

"local", i.e. driven by the local CDW strain, which is greatest at the contact. In addition, except for low voltages at which the $\tau_0$-distribution broadens, the resonant frequency is smaller away from the contact (i.e. the inertia is greater) than at the contact.

The "repolarization" times (and inertia) are much greater than those associated with the dielectric constant, i.e. small amplitude oscillations of the CDW about its pinned position (at T ~ 80 K, the dielectric time constant ~ 1 μs[17]), because polarization and repolarization involve large, nonlocal changes in the CDW phase.[7,8] On the other hand, the repolarization times shown in Figure 3 are much smaller than the time constants governing how the CDW responds to elastic strains of the crystal; the elastic time constants were measured to be ~ 10 ms at 2 $V_T$ and diverged at threshold.[18] This difference might seem surprising, since both the repolarization and "elastic" time constants are related to long length scale deformations of the CDW. That the elastic time constant is greater therefore suggests that changes in the elastic constants caused by CDW depinning, believed to be due to the effect of the strain on the optimum local CDW phase [18,19], require changes of the CDW on an even longer length scale than repolarization. (Another possible reason for the difference is that the elastic measurements[18] averaged over the entire sample length, whereas we measured the reflectance changes near the contacts, where repolarization is fastest.)

Figure 4 shows direct time traces, measured with a digital oscilloscope, of the change in reflectance of sample 1 in response to applied square wave voltages. As for Figure 3, measurements were made for light adjacent to the contact (center panels) and 100 μm positions (bottom panels). Each trace in the figure represents the average of 500 time sweeps, with time normalized to the square wave period. Also shown (the top panels) is the resulting current; the current overshoot, due to CDW repolarization after reversing the voltage,[1,20] is shown by arrows in the current traces. Figure 4a shows the responses to 253 Hz square-waves of different magnitudes. The decrease in the relaxation time with increasing voltage is clear in the ΔR traces, but in all cases this time is much longer than the "current overshoot" time; in fact, as has been pointed out in References [3,9], the current estimated from the overshoot is only a small fraction of the repolarization current.

The center panel of Figure 4a shows that, when the current changes from negative to positive, the reflectance adjacent to the contact has a delay of ~ 100 μs before it starts changing abruptly. This is seen more clearly in Figure 4b, at which the responses to 200 mV square waves at different frequencies are shown. Note that away from the contact (lower panels), while there is still a comparable delay in ΔR, the onset of the increase in reflectance is much broader (i.e. less abrupt). Also note that the delay is several times shorter when the current is changing from positive to negative. While the magnitude of the CDW polarization has previously been found to be polarity dependent,[21] for example the magnitude of the polarization has been observed to be larger on the negative side of the sample in $TaS_3$ at lower temperatures,[22] this is the first report of different dynamic response for the two polarities. This dynamic polarity dependence suggests that Schottky barriers at the contacts may contribute to the surprisingly large inertia. If this is so, perhaps different barriers on the two contacts are responsible for the very asymmetric spatial dependence. However, since the penetration depth of light at these frequencies is ~ 1000 A°, such a strong contact dependence is surprising. In future work, we will investigate samples with different contacts.

In fact, the delayed response of the reflectance as compared to the current suggests that our underlying assumption, that $\Delta R/R \propto \partial\varphi/\partial z$, may require qualification, since the current overshoot indicates that the CDW polarization is "undelayed". The difference may reflect the fact that the reflectance and current are probing different fractions of the sample depth. (Similarly, as mentioned above, differences in the spatial dependence of the changes in transmittance and reflectance were previously noted in blue bronze.[11]) It will therefore also be interesting to compare the time dependence of changes in transmittance with those in reflectance.

To some extent, the abrupt but delayed change in reflectance adjacent to the contact resembles the delays in CDW conduction, typically ~ 1 ms, that have been observed in samples exhibiting "switching", i.e. hysteretic discontinuous changes in current,[23, 24] but the resemblance may be superficial. The switching delays were associated with the time needed for the CDW polarization to build up to the value needed for phase slip,[24] but in our case, with non-switching samples, no delay was observed in the current change (including the overshoot).

Our measurements suggest that the electro-optic response can be used to measure the temporal/spatial dependence of changes in CDW polarization without adding multiple contacts, which tend to strongly perturb the local fields, especially in semiconducting CDW materials such as $TaS_3$ and blue bronze. Furthermore, once the spectrum is known, the variation of the polarization with depth in the sample can be measured by comparison of the response at different wavelengths and polarizations or, as mentioned above, by comparison of the electro-reflectance and electro-transmittance. However, we note two caveats: i) In all the measurements discussed above, the light spot filled the sample width, so that we averaged over this transverse dimension. In cases when a more narrow light spot was used, we frequently observed large differences in ΔR as the spot was scanned across the sample width, in some cases even changing sign. Similarly, large local changes in the chemical potential, implying local polarizations, varying across the sample width were measured with a scanning electron microscope,[25] presumably reflecting the sensitivity of the CDW to grain boundaries and surface features prevalent in even visually smooth samples of $TaS_3$. ii) For some cases, the measured frequency dependence of ΔR/R did not obey Eqtn. (2), which was introduced above simply to parameterize results for the two samples studied most extensively. In particular, in some cases high frequency inversion was not observed (i.e. $\Omega_0/2\pi > 4$ kHz) and in others the peak in the quadrature component of ΔR/R was much less symmetric than given by this equation.

In summary, we have presented results of measurements of the voltage, frequency, and spatial dependence of the electro-reflectance of orthorhombic $TaS_3$. The response indicates that there is considerable inertia, which is observable as a polarity dependent delay in ΔR in measurements of its time dependence. We have fit the frequency dependence of ΔR to a modified damped harmonic oscillator response, and find that the relaxation time varies as $\tau_0 \sim V^{-p}$, with p between 1 and 1.5, that the distribution of relaxation times increases at low voltages, and that the relaxation time increases away from the current contacts.

Crystals were provided by R.E. Thorne of Cornell University. We also appreciate helpful discussions with V.A. Bondarenko, P. Monceau and M.E. Itkis. This research was supported by the National Science Foundation, Grant DMR-0100572.

# **Figure Captions**

Figure 1. Spatial dependence of the in-phase (upper panel) and quadrature (lower panel) response of the reflectance (R) to 253 Hz square waves of magnitudes 60, 95, and 150 mV for sample 1 with light of energy $\nu = 860$ cm$^{-1}$. The right (x ~ 800 µm) contact is grounded and the voltage applied to the left (x ~ 0) contact, so that (noting the minus sign on the ordinate) the reflectance decreases for large voltages on the side of the sample which is positive. The inset shows the voltage dependence of the reflectance at the left contact.

Figure 2. Frequency dependence of the in-phase (upper panel) and quadrature (lower panel) electro-reflectance for 65 mV, 95 mV, 150 mV, and 235 mV square waves for sample 1, with light ($\nu = 860$ cm$^{-1}$) incident adjacent to the left contact (see Figure 1). The curves show fits to Eqtn. 2.

Figure 3. Voltage dependence of the parameters of Eqtn. 2. Closed symbols: sample 1 (measured at 860 cm$^{-1}$); open symbols: sample 2 (measured at 695 cm$^{-1}$). Circles: light focused adjacent to current contact; triangles: light focused 100 µm from this contact. The dotted lines show the onset voltages ($V_{onset} \approx 40$ mV for both samples) and thresholds for non-linear current ($V_T \approx 70$ mV for sample 1 and 100 mV for sample 2). The dashed line in the upper panel indicates $\tau_0 \, \alpha \, V^{-3/2}$ behavior for reference. (The symbol with the arrow indicates a fit for which the resonant frequency was effectively infinite, i.e. >> 4 kHz.)

Figure 4. Time dependence of the CDW current (top panels), reflectance adjacent to current contacts (center panels), and reflectance 100 μm from the contacts (bottom panels) for sample 1 at $\nu = 860$ cm$^{-1}$. The horizontal arrows in the top panels indicate current overshoot. a) Responses to $\Omega/2\pi = 253$ Hz square waves with magnitudes 110 mV, 150 mV, 300 mV. b) Responses to 200 mV square waves of frequencies $\Omega/2\pi = 253$ Hz, 714 Hz, and 2011 Hz.

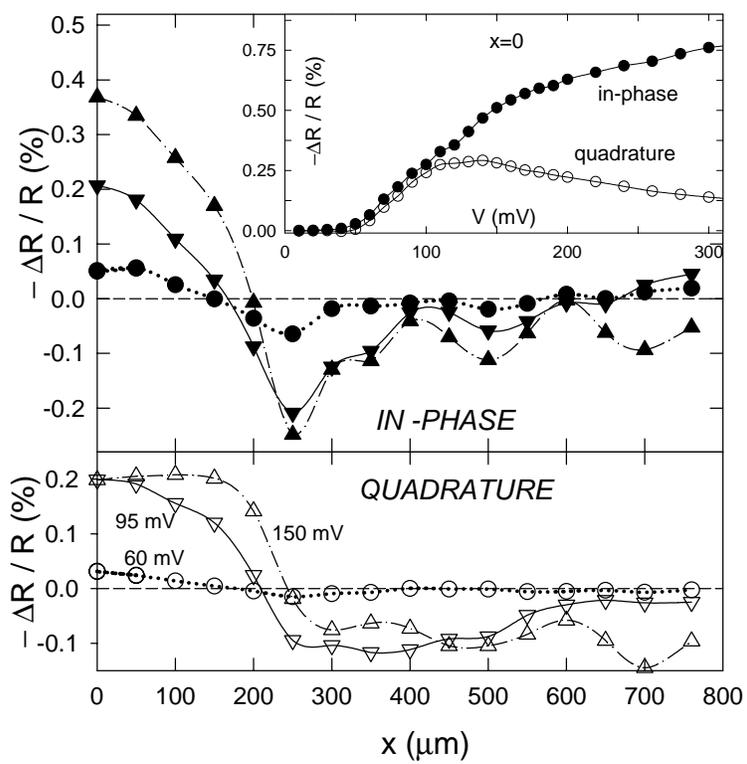

Figure 1.

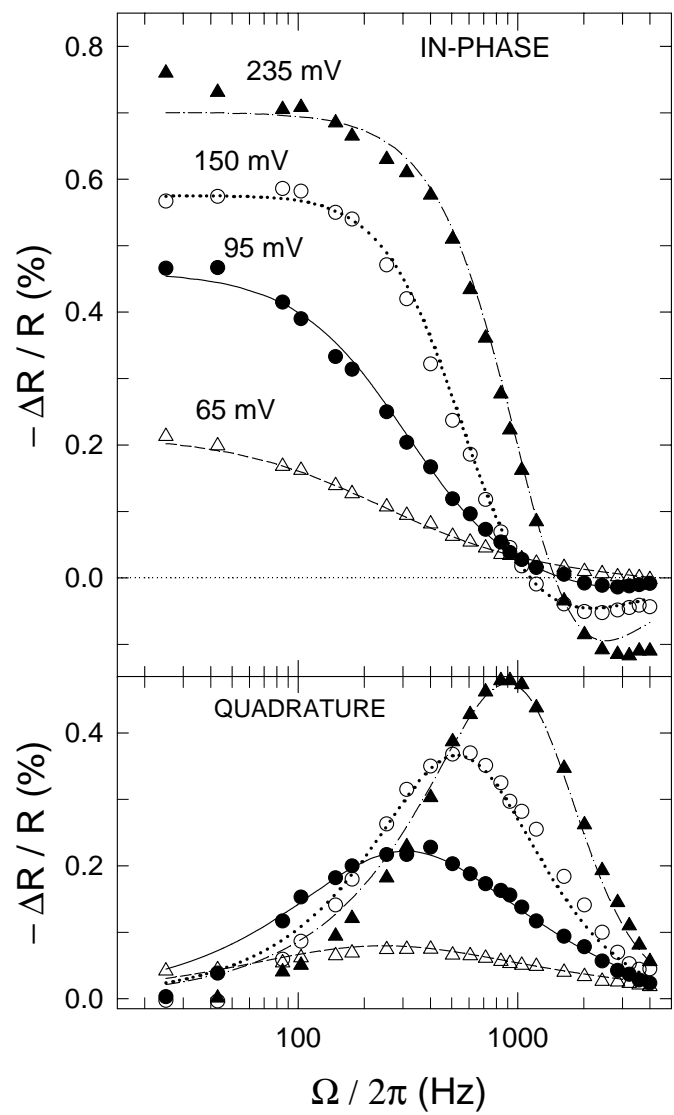

Figure 2.

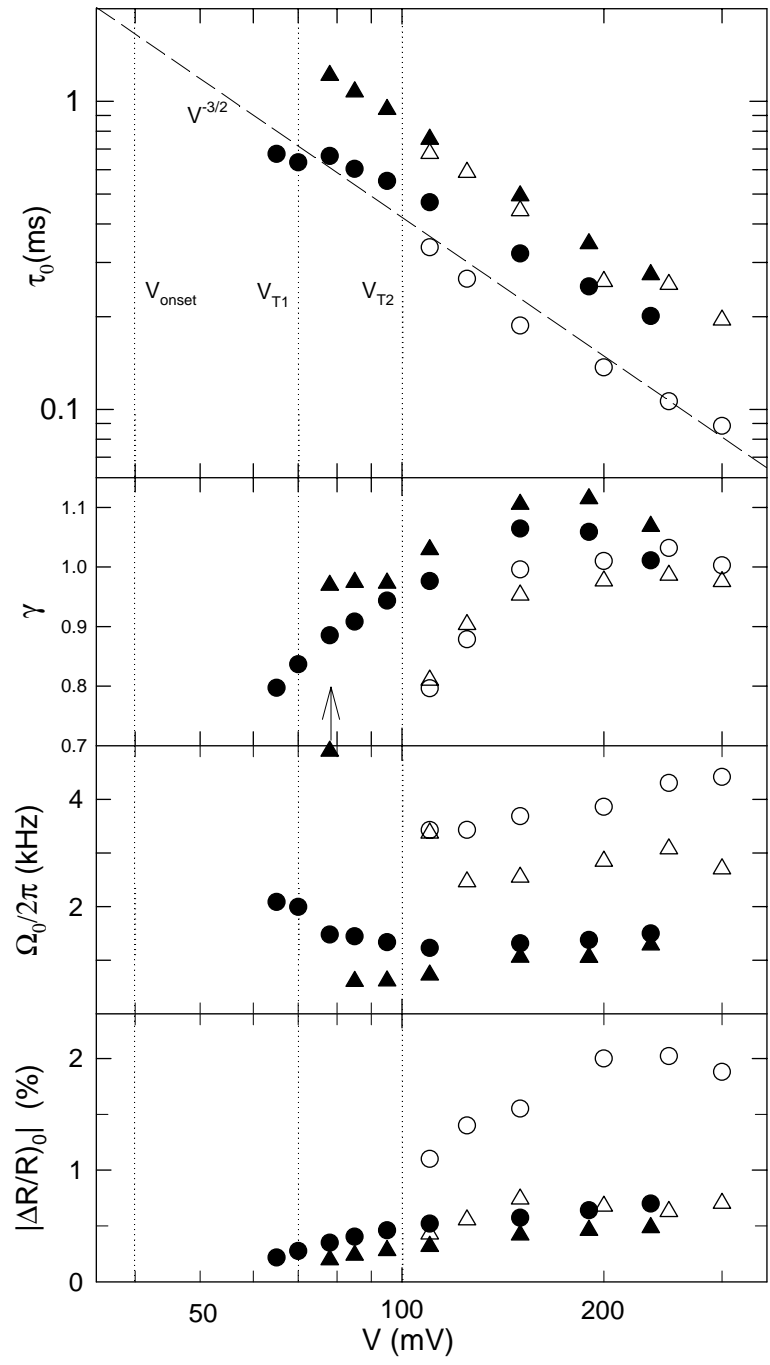

Figure 3

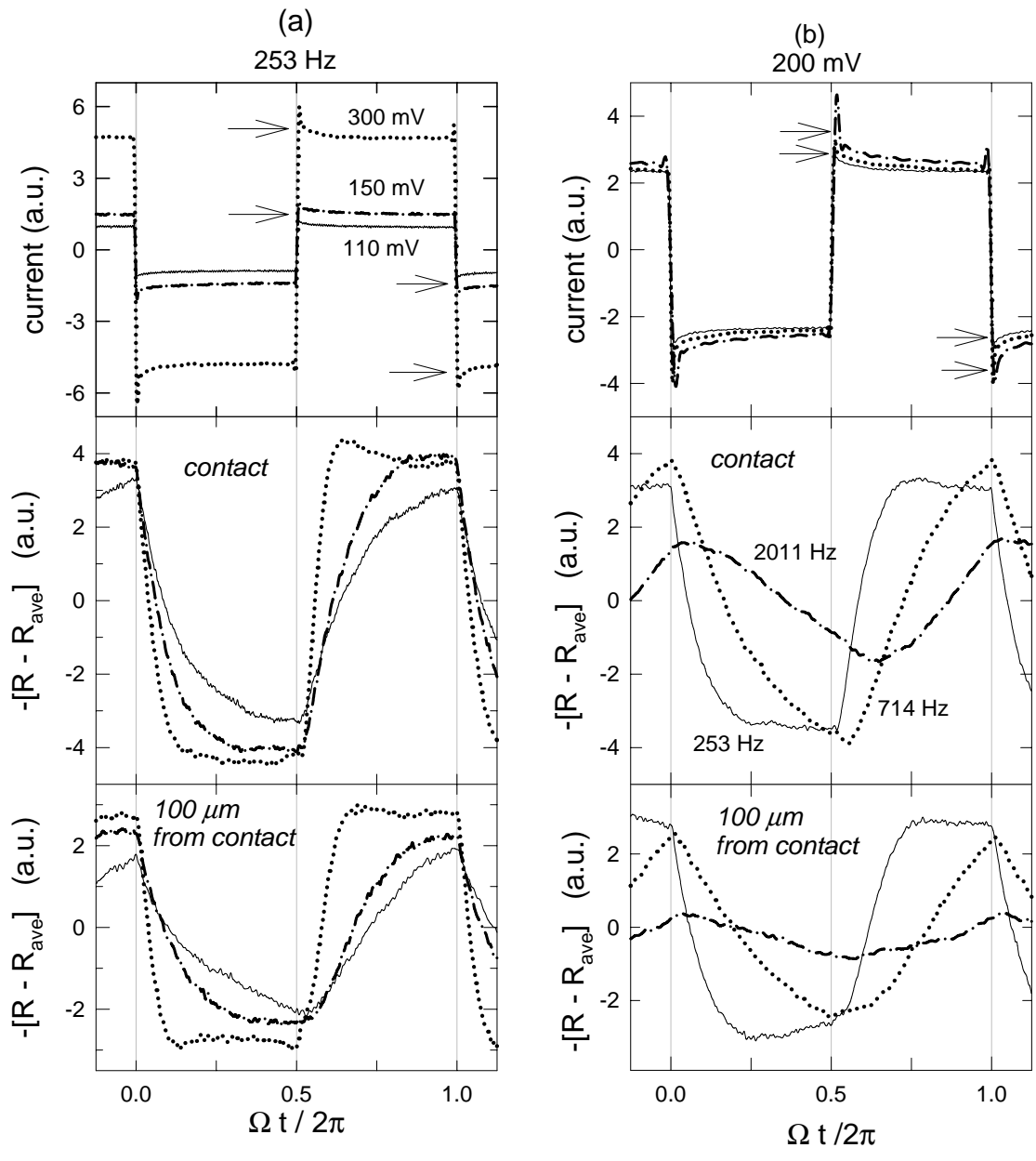

Figure 4